\documentclass[a4 paper, 12 pt] {article}
\def\be{\begin{equation}}
\def\ee{\end{equation}}
\def\bdi{\begin{displaymath}}
\def\edi{\end{displaymath}}
\def\br{\begin{eqnarray}}
\def\er{\end{eqnarray}}
\def\no{\nonumber}

\def\u2{\mid u\mid^2}

\def\no{\nonumber}

\def\RR{{\rm I\kern-.1567em R}}                              % Doppel R
 \def\CC{{\rm C\kern-4.7pt                                    % Doppel C
 \vrule height 7.7pt width 0.4pt depth -0.5pt \phantom {.}}}
 \def\ZZ{{\sf Z\kern-4.5pt Z}}

\begin{document}

\title{\bf{Hopf solitons and Hopf $Q$-balls on $S^3$}}

\author{C. Adam$^{a)}$\thanks{adam@fpaxp1.usc.es}, \, 
J. S\'{a}nchez-Guill\'{e}n$^{a)}$\thanks{joaquin@fpaxp1.usc.es} \, 
and  A. Wereszczy\'{n}ski$^{b)}$\thanks{wereszczynski@th.if.uj.edu.pl}
      \\ \\
       \small{ $^{a)}$Departamento de Fisica de Particulas, Universidad
       de Santiago}
       \\
       \small{ and Instituto Galego de Fisica de Altas Enerxias (IGFAE)}
       \\ \small{E-15782 Santiago de Compostela, Spain}
       \\ \small{ $^{b)}$Institute of Physics,  Jagiellonian University,}
       \\ \small{ Reymonta 4, 30-059 Krak\'{o}w, Poland} }

\maketitle

\begin{abstract}
Field theories with a $S^2$-valued unit vector field living on
$S^3 \times \RR$ space-time are investigated. The corresponding
eikonal equation, which is known to provide an integrable sector
for various sigma models in different spaces, is solved giving
static as well as time-dependent multiply knotted configurations
on $S^3$ with arbitrary values of the Hopf index. Using these
results, we then find a set of hopfions with topological charge
$Q_H=m^2$, $m \in \mathbf{Z}$, in the integrable subsector of the
pure $CP^1$ model. In addition, we show that the $CP^1$ model with
a potential term provides time-dependent solitons. In the case of
the so-called "new baby Skyrme" potential we find, e.g., exact
stationary hopfions, i.e., topological $Q$-balls. \\
Our results further enable us to construct exact static and
stationary Hopf solitons in the Faddeev--Niemi model with or
without the new baby Skyrme potential. Generalizations for a large
class of models are also discussed.
\end{abstract}
\newpage
%%%%%%%%%%%%%%%%%%%%%%%%%%%%%%%%%%%%%%%%%%%%%%%%%%%%%%%%%%%%%%
\section{Introduction}
%%%%%%%%%%%%%%%%%%%%%%%%%%%%%%%%%%%%%%%%%%%%%%%%%%%%%%%%%%%%%%%
Dynamical models allowing for stable knot-like structures seem to
play an increasingly important role in modern physics. For
instance, knotted solitons find some applications in condensed
matter physics \cite{babaev}, \cite{cho4} as topological defects in
multi-component Bose condensates. On the other hand, in high
energy physics the rising interest originates in the idea that
glueballs, i.e., effective particle-like excitations in the low
energy limit of quantum gluodynamics, may be understood as closed,
in general knotted, tubes of the squeezed color field (possibly
due to the dual Meissner effect \cite{t'hooft}). In
fact, such a framework is in  accordance with the standard picture
of mesons where a quark and an antiquark are connected by a thin
flux-tube of the gauge field. When the quark sources are absent,
the ends of the tube must join to form a (in general knotted)
loop. There has been made much effort to derive such a qualitative
picture from the original quantum theory and to find the correct
low energy effective action with knotted solitons as stable
excitations. One well known proposal is the Faddeev--Niemi model
\cite{niemi1}, which is, in fact, just the $S^2$ restriction of
the Skyrme model, as can be explicitly demonstrated, see \cite{cho3},
\begin{equation}
{\cal L}_{FN}=
\frac{1}{2}\mu^2 {\cal L}_2 -\frac{1}{4e^2} {\cal L}_4, \label{fn stand}
\end{equation}
where 
\be {\cal L}_2 \equiv (\partial_{\mu} \vec{n})^2 = 4
\frac{\partial^\mu u \partial_\mu \bar u}{(1 + u\bar u)^2} , 
\ee
\be 
{\cal L}_4 \equiv [\vec{n} \cdot (\partial_{\mu} \vec{n}
\times
\partial_{\nu} \vec{n})]^2 =
8 \frac{(\partial^\mu u \partial_\mu \bar u)^2 - (\partial^\mu u
\partial_\mu u)(\partial^\nu \bar u \partial_\nu \bar u)}{(1+u\bar u)^4} ,
\ee 
where $\mu$ is a constant with the dimension of a mass, and
$e$ is a dimensionless constant. Further, $\vec{n}$ is a real
three component unit vector field living in $(3+1)$ Minkowski
space-time, and $u$ is a complex scalar field  related to the unit
vector field by the standard stereographic projection
\begin{equation}
\vec{n}= \frac{1}{1+|u|^2} ( u+\bar{u}, -i(u-\bar{u}), |u|^2-1).
\label{stereograf}
\end{equation}
There are some arguments that this field, connected with the
primary gauge field via the Cho-Faddeev-Niemi decompositions,
might describe the infrared relevant degrees of freedom of quantum
gluodynamics \cite{niemi2}, \cite{cho}, \cite{shabanov}, \cite{kondo},
\cite{cho2}. 
On the other hand, the stability of the spectrum and
even of the field decomposition under quantum fluctuations is a
matter of active research and discussion \cite{Wipf2},
\cite{Gies1}.
\\
It has been proved that the Faddeev--Niemi model indeed supports
knotted solitons \cite{LiYa} with a nonzero value of the Hopf
index $Q_H \in \pi_3(S^2)$. However, only numerical solutions have
been reported \cite{battyde}, \cite{salo} and many important
questions concerning, e.g.,  the geometry of the stable (or
meta-stable) configurations in a fixed topological sector are
still unsolved.
\\
In order to understand the behavior of hopfions in an analytical
way and to test some ideas borrowed from other soliton systems,
two dynamical models have been proposed. They are known as the
Nicole \cite{nicole}
\begin{equation}
{\cal L}_{Ni}=\frac{1}{2} {\cal L}_2^{\frac{3}{2}} \label{nicole
stand}
\end{equation}
and Aratyn-Ferreira-Zimerman model \cite{aratyn}
\begin{equation}
{\cal L}_{AFZ}= \frac{1}{4} {\cal L}_4^{\frac{3}{4}}.
\label{arratyn stand}
\end{equation}
A common feature of these two nonlinear models is their invariance
under scale reparametrizations. This provides a new way
(originally proposed by Deser et. al. \cite{deser}) to circumvent
Derrick's theorem. In addition, they possess exact soliton
solutions.
\\
However, there is a different strategy to construct exact
hopfions. Namely, it is possible to change the base space in such
a manner that the topological content of the theory remains
unchanged. The most obvious proposition is to investigate fields
on a three dimensional sphere $S^3_{R_0}$, where $R_0$ is its
radius, instead of the standard three dimensional Euclidean space
$\RR^3$ \cite{ward}, \cite{ferreira}. In this case, the
introduction of the new parameter $R_0$ sets the scale in the
model and therefore gives an alternative way to circumvent
Derrick's theorem about the nonexistence of static solitons. Hopf
solitons on $S^3$ have been recently considered by R. Ward and L.
A. Ferreira {\it et. al.} in the context of Faddeev--Niemi
\cite{ward} and AFZ-like models \cite{ferreira}, respectively. In
the present paper, we would like to further develop these
investigations. Concretely, in Section 2 we construct a family of
solutions for the static as well as for the time-dependent eikonal
equation for arbitrary values of the Hopf index. The eikonal
equation defines integrable subsectors for the models discussed in
the subsequent sections and its solutions, therefore, will help us
in finding explicit soliton solutions. In Section 3 we study the
$CP(1)$ model and find a class of static solutions and, when a
potential is added, a similar class of time-dependent solutions.
The same results can be found for a family of generalized $CP(1)$
models. In Section 4 we study another family of models, among
which the Faddeev--Niemi model can be found. We establish the
existence of a Hopf soliton with topological charge one for all of
them. Further, for the Faddeev--Niemi model with and without a
potential term, the existence of a time-dependent, stationary
solution is demonstrated. In addition, we construct some
generalizations of these models, which have solitons with higher
Hopf index. Section 5 contains our conclusions.
\\
To finish the introduction, let us remind some details of the
geometry of the three-sphere. A three-sphere $S^3$ with radius
$R_0$ embedded in four-dimensional Euclidean space $\RR^4$ is
described by the equation \be \label{Surf-S3} X_1^2 +X_2^2 +X_3^2
+X_4^2 =R_0^2 \ee where the $X_i$ are the usual orthonormal
coordinates in $\RR^4$. Further, the metric on the surface defined
by Eq. (\ref{Surf-S3}) is induced by the standard Euclidean metric
on $\RR^4$. Introducing coordinates on $S^3$ as in
\cite{ferreira}, \br X_1 =R_0 \sqrt{z} \cos \phi_2 & ,& X_3 = R_0
\sqrt{1-z} \cos \phi_1 \no
\\
X_2 =R_0 \sqrt{z} \sin \phi_2 & ,& X_4 = R_0 \sqrt{1-z} \sin
\phi_1 \er where $z \in [0,1]$ and the angles $\phi_1, \phi_2 \in
[0,2\pi]$, the metric and volume form on space-time $\RR \times
S^3$ are
\begin{equation}
ds^2 = dt^2 -R_0^2\left( \frac{dz^2}{4z(1-z)} +(1-z)d\phi_1^2
+zd\phi_2^2 \right) , \label{metric}
\end{equation}
\be dV = \frac{1}{2} dtdzd\phi_1 d\phi_2 . \ee
%%%%%%%%%%%%%%%%%%%%%%%%%%%%%%%%%%%%%%%%%%%%%%%%%%%%%%%%%%%%%%%
\section{Eikonal knots on $ S^3 $}
%%%%%%%%%%%%%%%%%%%%%%%%%%%%%%%%%%%%%%%%%%%%%%%%%%%%%%%%%%%%%%%
The construction of soliton solutions with non-zero Hopf index is
sometimes facilitated by  restricting  the original theory to an
integrable submodel \cite{integrability}, where integrability is
understood as the existence of an infinite number of local
conserved currents. In typical situations (Nicole or Faddeev--Niemi
model), such an integrable subsystem can be defined by imposing
the complex eikonal equation \cite{eikonal}\footnote{For some
Lagrangian-dependent "generalizations" of the eikonal equation and
their application to integrability see \cite{general eikonal}.
Moreover, a weaker integrability condition has been investigated
in \cite{weak}.}
\begin{equation}
(\partial_{\mu} u)^2=0. \label{eikonal eq def}
\end{equation}
Solutions of the complex eikonal equation on $\RR^3$ describe
(linked) torus knots with an arbitrary value of the Hopf charge
\cite{adam1}, \cite{were1} and in some particular cases may help
to derive hopfions in dynamical systems \cite{were2}. On the other
hand, the fact that the eikonal equation is an integrability
condition for sigma models does not depend on the base space.
Thus, as our aim is to study knotted configurations on $S^3$, it
is important to solve the eikonal on the three-sphere as well. In
fact, all solitons which we study will obey the eikonal equation,
as well.
\\
Let us assume the following static Ansatz \cite{aratyn},
\cite{ferreira}
\begin{equation}
u_0=f(z)e^{i(m_1\phi_1+m_2\phi_2)}, \label{anzatz}
\end{equation}
where $m_1, m_2$ are integer numbers. Then we find
\begin{equation}
\nabla u =\frac{1}{R_0} \left[ 2\sqrt{z(1-z)} f' \hat{e}_z +
\frac{im_1f}{\sqrt{1-z}} \hat{e}_{\phi_1} + \frac{im_2f}{\sqrt{z}}
\hat{e}_{\phi_2} \right] e^{i(m_1\phi_1+m_2\phi_2)} \label{nabla}
\end{equation}
and equation (\ref{eikonal eq def}) can be rewritten as follows
\begin{equation}
4z(1-z)f'^2 - \frac{f^2}{z(1-z)} \left( zm_1^2+(1-z)m_2^2 \right)
=0 \label{eikonal eq1}
\end{equation}
or
\begin{equation}
\frac{f'^2}{f^2} = \frac{zm_1^2+(1-z)m_2^2}{4z^2(1-z)^2}.
\label{eikonal eq2}
\end{equation}
One can solve this equation and obtain the following solutions
\begin{equation}
f_\pm = C \left[ \left(
\frac{m_1+\sqrt{(m_1^2-m_2^2)z+m_2^2}}{m_1-\sqrt{(m_1^2-m_2^2)z+m_2^2}}
\right)^{m_1} \left( \frac{-m_2+\sqrt{(m_1^2-m_2^2)z+m_2^2}}{m_2+
\sqrt{(m_1^2-m_2^2)z+m_2^2}}\right)^{m_2} \right]^{\pm
\frac{1}{2}}. \label{eikonal sol}
\end{equation}
Here $C$ is a complex constant. Our solutions simplify a lot if we
assume the special case $m_1=m_2=m$. Then
\begin{equation}
f_\pm = C \left( \frac{1}{z} -1 \right)^{\pm \frac{m}{2}}.
\label{eikonal sol sym}
\end{equation}
The point is that such profile functions, if inserted into the
Ansatz, give configurations with a non-trivial value of the
pertinent topological charge. Namely \cite{ferreira},
\begin{equation}
Q_H=\pm m_1 m_2. \label{eikonal charge}
\end{equation}
Moreover, taking into account symmetries of the complex eikonal
equation we can find more general solutions
\begin{equation}
u = F(u_0), \label{eikonal sol gen}
\end{equation}
where $F$ is any (anti)holomorphic function of the basic solution
$u_0$. Thus, we can conclude that the complex eikonal equation on
$S^3$ describes linked torus knots, as its counterpart on $R^3$.
\\
The eikonal equation on $S^3$ also enables us to obtain
time-dependent knotted configurations, unlike the standard $\RR^3$
case where no time-dependent solutions are known. The non-static
eikonal equation has the form
\begin{equation}
(\partial_t u )^2 -(\nabla u)^2=0. \label{time eikonal eq def}
\end{equation}
Let us assume that the time-dependence can be factorized. Due to
equation (\ref{time eikonal eq def}) such factorization must have
the form of an exponential
\begin{equation}
u_0=f(z)e^{\pm \lambda t} e^{i(m_1\phi_1+m_2\phi_2)}, \label{time
anzatz}
\end{equation}
where $\lambda$ is a complex parameter and $f(z)$ is a new profile
function yet to be determined. It is straightforward to notice
that there are two generic situations.
\\
First of all, for $\lambda \in R$ we can find exploding or
collapsing solutions, depending on the sign of the parameter. Now,
formula (\ref{time eikonal eq def}) takes the form
\begin{equation}
4z(1-z)f'^2 - f^2 \left[ \frac{zm_1^2+(1-z)m_2^2}{z(1-z)} + R_0^2
\lambda^2 \right] =0 \label{time eikonal eq1}
\end{equation}
or
\begin{equation}
\frac{f'^2}{f^2} = \frac{1}{4} \left[
\frac{zm_1^2+(1-z)m_2^2}{z^2(1-z)^2} + \frac{R_0^2\lambda^2
}{z(1-z)}  \right]. \label{time eikonal eq2}
\end{equation}
This equation can be integrated giving exact but rather
complicated solutions for the shape function
$$
f_\pm (z)=C \, \mbox{exp} \, \left[ \mp \frac{1}{2a} \arctan
\left( \frac{1+a^2(m_1^2-m_2^2)-2z}{2\sqrt{z(1-z) +a^2(m_2^2(1-z)
+m_1^2z}}\right) \right] \times
$$
$$
\left[ \frac{-z +a^2(m_2^2(-2+z)-m_1^2z) +2am_2 \sqrt{z(1-z)
+a^2(m_2^2(1-z) +m_1^2z)} }{ a^3m_2^3z} \right]^{\pm
\frac{m_2}{2}}
$$
\begin{equation}
\left[ \frac{1-z +a^2(m_2^2(1-z)-m_1^2(1+z)) -2am_1 \sqrt{z(1-z)
+a^2(m_2^2(1-z) +m_1^2z)} }{ a^3m_1^3(-1+z)} \right]^{\mp \frac{
m_1}{2}}, \label{time collaps sol1}
\end{equation}
where $C$ is a complex constant and
\begin{equation}
a^2=\frac{1}{R_0^2 \lambda^2}. \label{def a}
\end{equation}
It is easy to see that these new profile functions are
asymptotically (for $z \rightarrow 0$ and $ z \rightarrow 1$)
identical to their static counterparts. The additional term in the
eikonal equation only modifies the behavior in the intermediate
region. Therefore, the topological features of the time-dependent
solutions are analogous to the static case.
\\
Our solutions take a simpler form if $m_1=m_2=m$,
$$
f_\pm (z)=C \, \mbox{exp} \, \left[\mp \frac{1}{2a} \arctan \left(
\frac{1-2z}{2\sqrt{m^2a^2 +z-z^2}}\right)\right] \times $$
\begin{equation} \left(
\frac{1-z}{z} \cdot \frac{z+2ma(ma+\sqrt{m^2a^2+z-z^2})}{1-z+
2ma(ma+\sqrt{m^2a^2+z-z^2})} \right)^{\mp \frac{m}{2}}.
\label{time collaps sol2}
\end{equation}
Another type of time-dependent solutions is a family of
time-periodic configurations. Now $\lambda = i \omega $, where
$\omega \in R$. Thus,
\begin{equation}
u_0=f(z)e^{\pm i \omega t} e^{i(m_1\phi_1+m_2\phi_2)},
\label{periodic anzatz}
\end{equation}
where the unknown shape function satisfies the following equation
\begin{equation}
\frac{f'^2}{f^2} = \frac{1}{4} \left[
\frac{zm_1^2+(1-z)m_2^2}{z^2(1-z)^2} - \frac{R_0^2\omega^2
}{z(1-z)} \right]. \label{periodic eikonal eq1}
\end{equation}
In this case the solution is
$$ f_\pm (z)
=C \left(-1 +a^2 (m_1^2-m_2^2) +2z +2\sqrt{a^2(m_2^2 (1-z)
+m_1^2z) -z(1-z)} \right)^{\mp \frac{1}{2a}} \times $$
$$ \left( \frac{-1+z+a^2(m_2^2(1-z)+m_1^2(1+z)) +2\sqrt{a^2(m_2^2 (1-z)
+m_1^2z) -z(1-z)}}{a^3m_1^3(-1+z)}\right)^{\pm \frac{m_1}{2}} $$
\begin{equation}
\left( \frac{z+a^2(m_2^2(-2+z)-m_1^2z) -2\sqrt{a^2(m_2^2 (1-z)
+m_1^2z) -z(1-z)}}{a^3m_2^3z}\right)^{\mp \frac{m_2}{2}}
\label{time periodic sol1}
\end{equation}
or in the simpler case, when $m_1=m_2=m$,
$$ f_\pm (z)= C \left( -1+2z +2\sqrt{ m^2a^2-z(1-z)}\right)^{\pm
\frac{1}{2a}} \times $$
\begin{equation}
\left( \frac{-1+z}{z} \cdot \frac{z-2ma(ma+\sqrt{
m^2a^2-z(1-z)})}{-1+z+2ma(ma+\sqrt{ ma^2-z(1-z)})} \right)^{\pm
\frac{m}{2}}
\end{equation}
Such shape functions give non-trivial topological configurations
if $f$ is a smooth, real function which tends to $0$ for $z
\rightarrow 0$ and to $\infty$ when $z \rightarrow 1$ (or
inversely). Therefore we get a restriction for the frequencies of
the stationary solutions with a fixed topological charge
\begin{equation}
\omega^2 \leq \frac{2(m^2_1+m_2^2)}{R_0^2}. \label{freq rel}
\end{equation}
In other words, there is an upper bound for the frequencies of a
stationary solution. Only configurations with lower frequencies
can be constructed. Eq. (\ref{freq rel}) leads to two important
observations. Firstly, the range of possible frequencies grows
with the topological charge. Knots with higher topological charge
can have higher frequencies. Secondly, the range becomes narrower
if the radius of the base space grows. Thus, for the Euclidean
space, i.e., when $R \rightarrow \infty$, no time-periodic eikonal
knots can be found.
\\
At the end of this section let us notice that the complex eikonal
equation admits also topologically trivial solutions. An
interesting example can be found if we assume that the complex
field is a function only of time and $z$ variable. Then using the
method of characteristics we derive the following general solution
\begin{equation}
u=u(t \pm \frac{1}{2} \arcsin (1-2z)). \label{plane wave}
\end{equation}
One can immediately see that such a solution describes a very
nonlinear travelling wave.
%%%%%%%%%%%%%%%%%%%%%%%%%%%%%%%%%%%%%%%%%%%%%%%%%%%%%%%%%%%%%%%
\section{$ CP^1 $ model on $S^3$}
%%%%%%%%%%%%%%%%%%%%%%%%%%%%%%%%%%%%%%%%%%%%%%%%%%%%%%%%%%%%%%%
So far, the considered knots have been only solutions of the
complex eikonal equation, without any underlying Lagrange
structure. In the next sections we show that at least some of the
eikonal knots appear as  solutions of a large family of
nonlinear sigma models on $S^3$. Let us mention at this point that
for the purely quartic model with Lagrangian ${\cal L}_4$ - which
is integrable without any additional constraint - both static and
time-dependent, stationary hopfions on $S^3$ have been found and
studied in Ref. \cite{ferreira}.
%%%%%%%%%%%%%%%%%%%%%%%%%%%%%%%%%%%%%%%%%%%%%%%%%%%%%%%%%%%%%%%
\subsection{Static solitons}
%%%%%%%%%%%%%%%%%%%%%%%%%%%%%%%%%%%%%%%%%%%%%%%%%%%%%%%%%%%%%%%
Let us start with the simplest example, i.e., the $CP^1$ model
\begin{equation}
\mathcal{L}_{CP^1} \equiv \frac{1}{4} {\cal L}_2 =
\frac{\partial_{\mu} u \partial^{\mu} \bar{u}}{(1+|u|^2)^2}.
\label{cp1}
\end{equation}
The equation of motion reads
\begin{equation}
\partial^2_{\mu} u - \frac{2\bar{u}}{1+|u|^2}
(\partial_{\mu} u )^2=0. \label{cp1 eq mot2}
\end{equation}
This equation is certainly satisfied for a submodel, where the
complex field $u$ obeys the two equations
\begin{equation}
\partial^2_{\mu} u =0 \; \; \; \mbox{and} \; \; \; (\partial_{\mu} u
)^2=0 , \label{cp1 submodel}
\end{equation}
i.e.,  the wave equation and  the eikonal equation.\footnote{This
pair of equations has been studied first, in the context of the
$CP^1$ model in $2+1$ dimensional space-time, in Ref.
\cite{Ward1}.} Due to the fact that the eikonal equation is
imposed, this submodel belongs to the integrable systems.
\\
In order to find static knotted solitons we assume the same ansatz
as in (\ref{anzatz}). Then
\begin{equation}
\nabla^2 u = \frac{1}{R_0^2} \left[ 4\partial_z \left( z(1-z) f'
\right) -f \left( \frac{m_1^2}{1-z} + \frac{m_2^2}{z}
\right)\right] e^{i(m_1\phi_1+m_2\phi_2)} \label{laplace}
\end{equation}
and the first equation in (\ref{cp1 submodel}) can be rewritten as
\begin{equation}
4\partial_z \left( z(1-z) f' \right) = f \left( \frac{m_1^2}{1-z}
+ \frac{m_2^2}{z} \right). \label{laplace eq1}
\end{equation}
Of course, as the subsystem consists of the static eikonal
equation, as well, the profile function $f$ has to satisfy
equation (\ref{eikonal eq2}). Therefore, the left hand side of
({\ref{laplace eq1}) can be expressed by (\ref{eikonal eq2}). Then
we obtain
\begin{equation}
4 \partial_z \left( z(1-z)f' \right) = 4z(1-z) f'\frac{f'}{f}.
\label{submodel cp1 eq1}
\end{equation}
A first integration leads to
\begin{equation}
\ln \left( \frac{1}{b} \frac{z(1-z)f'}{f} \right) =0,
\label{submodel cp1 eq2}
\end{equation}
which possesses the following solutions
\begin{equation}
f=B \left( \frac{1}{z} -1 \right)^{b}. \label{cp1 shape sol}
\end{equation}
Here $B$ and $b$ are arbitrary, in general complex constants.
However, as this solution should satisfied also the eikonal
equation, we find that $b=\pm m/2$. Therefore, the field
configurations
\begin{equation}
u_\pm =B \left( \frac{1}{z} -1 \right)^{\pm \frac{m}{2}}
e^{im(\phi_1+\phi_2)} \label{cp1 sol}
\end{equation}
are solutions of the submodel (\ref{cp1 submodel}) and, as a
consequence, they are static solutions of $CP^1$ model. Moreover,
they carry nonzero value of the Hopf index
\begin{equation}
Q_H=\pm m^2. \label{cp1 charge}
\end{equation}
\\
More complicated solutions can be constructed if we take advantage
of the symmetries of the submodel. In fact, it is easily checked
that any $\tilde u$ of the form
\begin{equation}
\tilde{u} = F(u), \label{cp1 sol gen}
\end{equation}
is a solution, where $F$ is any (anti)holomorphic function, and
$u$ ($\bar u$) is a solution of the submodel (e.g., of the form
(\ref{cp1 sol}) derived above). Thus, we can obtain quite
complicated linked  configurations with arbitrary Hopf charge.
\\
Let us now calculate the energies of the obtained solutions. They
are given by
\begin{equation}
E=\int_{S^3}  \frac{\nabla u \nabla \bar{u} }{(1+|u|^2)^2}
\frac{1}{2} dzd\phi_1 d\phi_2. \label{cp1 energy1}
\end{equation}
Thus, inserting (\ref{cp1 sol}) we get
\begin{equation}
E= 4\pi^2 m^2 R_0 \int_0^1 \frac{1}{z(1-z)} \frac{f^2}{(1+f^2)^2}
dz \label{cp1 energy2}
\end{equation}
and finally
\begin{equation}
E= 4\pi^2 R_0 |m|. \label{cp1 energy sol}
\end{equation}
Re-introducing the Hopf index we therefore find
\begin{equation}
E= 4\pi^2 R_0 |Q_H|^{\frac{1}{2}} , \label{cp1 energy charge}
\end{equation}
i.e., the energies grow like the square root of the Hopf index.
One can also observe that there is some degeneracy in the
energy spectrum, because the energy remains the same for all values of
the parameter $B$. As we will discuss in the last section, these solutions
are not stable, but rather saddle point solutions.
%%%%%%%%%%%%%%%%%%%%%%%%%%%%%%%%%%%%%%%%%%%%%%%%%%%%%%%%%%%%%%%
\subsection{Time-dependent solutions}
%%%%%%%%%%%%%%%%%%%%%%%%%%%%%%%%%%%%%%%%%%%%%%%%%%%%%%%%%%%%%%%
Exact, time-depended hopfions can be found if we consider the
$CP^1$ model with a potential explicitly breaking the global
$O(3)$ symmetry
\begin{equation}
\mathcal{L}=\frac{1}{4} {\cal L}_2 - V_{\rm I} \equiv
\frac{\partial_{\mu} u \partial^{\mu} \bar{u}}{(1+|u|^2)^2} -
\frac{\beta^2}{4} \frac{|u|^2}{(1+|u|^2)^2}. \label{dyn model}
\end{equation}
Here $\beta^2 $ is a positive constant. Such a potential has a
very simple form if we express it in terms of the original unit
vector field
\begin{equation}
V_{\rm I} (\vec{n})=\frac{\beta^2}{16} \left[ 1-(n^3)^2 \right].
\label{periodic potential}
\end{equation}
It is worth mentioning that this potential has been previously
considered in the context of the Skyrme model on the plane. More
precisely, it is the potential part of the so-called "new baby
Skyrme model" \cite{new baby}, which stabilizes the topological
solitons in that model. In addition, possible applications of
sigma model type theories to the low-energy sector of YM theory
require the explicit breaking of the global $O(3)$ symmetry, which
may be achieved, e.g., by the introduction of a symmetry-breaking
potential like the one chosen above \cite{FaNi3}, \cite{wipf} (see
also our remark in the summary section).
\\
The equation of motion for our model is
\begin{equation}
\frac{1}{(1+|u|^2)^2} \partial_{\mu} \partial^{\mu}u -
\frac{2\bar{u}}{(1+|u|^2)^3} (\partial u)^2 +\frac{\beta^2}{4}
\frac{u}{(1+|u|^2)^2} -\frac{\beta^2 u |u|^2}{2(1+|u|^2)^3} =0.
\label{dyn eqmot1}
\end{equation}
Similarly as in the pure $CP^1$ model it is possible to define a
submodel consisting of two, relative simple equations: a dynamical
one,
\begin{equation}
\partial_{\mu} \partial^{\mu}u + \frac{\beta^2}{4} u =0 \label{dyn
submodel1}
\end{equation}
and a constraint being a modification of the eikonal equation
\begin{equation}
2 \bar{u} (\partial_{\mu} u)^2 + \frac{\beta^2}{2} u |u|^2=0.
\label{dyn submodel2}
\end{equation}
Obviously, every solution of the subsystem obeys the equation of
motion for the full model. Notice that such a submodel is a
``massive'' modification of the pure $CP^1$ submodel with a
``imaginary mass''. In particular, formula (\ref{dyn submodel2})
can be rewritten in the form of the massive eikonal equation
\cite{were3}
\begin{equation}
(\partial_{\mu} u)^2 -M^2 u^2=0, \label{massive model}
\end{equation}
where the ``mass'' parameter $M^2 = - \beta^2/4$.
\\
Once again solutions of the submodel (\ref{dyn submodel1}),
(\ref{dyn submodel2}) are assumed in the form
\begin{equation}
u= f(z) e^{\pm i \omega t} e^{i(m_1\phi_1+m_2\phi_2)}. \label{dyn
anzatz}
\end{equation}
Then, we derive the following equations for the unknown shape
function
\begin{equation}
4\partial_z \left( z(1-z) f' \right) - f \left[ \frac{m_1}{1-z}
+\frac{m_2}{z} - R_0^2\omega^2 +\frac{\beta^2 R_0^2}{4} \right]=0,
\label{dyn submodel1 eq1}
\end{equation}
\begin{equation}
4z(1-z) f'^2-f^2\left[\frac{m_1}{1-z} +\frac{m_2}{z} -
R_0^2\omega^2 +\frac{\beta^2 R_0^2}{4} \right]=0.
\end{equation}
These expressions can be simplified if we impose an additional
condition for the frequency of the stationary solutions
\begin{equation}
\omega^2= \frac{\beta^2}{4}. \label{dynsol  cond}
\end{equation}
Then we get a set of equations which are identical to the static
equations in the pure $CP^1$ model. Therefore the shape function
is given by
\begin{equation}
f(z)=C \left( \frac{1}{z}-1 \right)^{\pm \frac{m}{2}},
\label{dynsol shape}
\end{equation}
where $m=m_1=m_2$ and $C$ is a complex constant. To summarize, we
have found a family of topologically nontrivial, stationary hopfions
\begin{equation}
u= C \left( \frac{1}{z}-1 \right)^{\pm \frac{m}{2}} e^{\pm i
\frac{\beta}{2} t} e^{im(\phi_1+\phi_2)}. \label{dyn sol}
\end{equation}
Such stationary configurations which, although they rotate in an
internal space, possess time independent energy density, are known
as $Q$-balls. They provide, e.g., a well known
example of nontopological solitons.
In the non-topological case these objects normally carry a
conserved charge where the conserved current is a Noether
current originating from an unbroken continuous global symmetry. In
our case it is the remaining unbroken $O(2)$ subgroup of $O(3)$.
Our solutions, however, have a conserved topological
charge in addition to the nontopological Noether charge
\begin{equation}
Q=i \int \frac{\bar{u} \partial_t u -u \partial_t
\bar{u}}{(1+|u|^2)^2} d V. \label{n charge}
\end{equation}
The energy of the stationary hopfions reads
\begin{equation}
E=\int dV \left( \frac{\nabla u \nabla \bar{u} }{(1+|u|^2)^2} +
\frac{\partial_t u \partial_t \bar{u} }{(1+|u|^2)^2} +
\frac{\beta^2}{4}  \frac{|u|^2}{(1+|u|^2)^2} \right) .
\label{q energy 1}
\end{equation}
Thus
\begin{equation}
E=4\pi^2 R_0|Q_H|^{\frac{1}{2}} + \frac{1}{2}|\beta Q|,
\label{q energy 2}
\end{equation}
where formula (\ref{cp1 energy sol}) has been taken into account.
As one might have expected, the $Q$-hopfions modify the standard $CP^1$
model in such a way that the degeneracy in the energy is lifted.
\\
It should be noticed that analogous stationary solutions of the
$CP^1$ model with the new baby Skyrme potential living in
(2+1) dimensional Min\-kow\-ski space-time and carrying the
pertinent topological charge (winding number) have been previously
found by Leese \cite{leese}. They are known as $Q$-lumps.
\\
Let us mention an interesting difference between $Q$-lumps
and $Q$-hopfions. $Q$-hopfions have finite Noether charge and
finite energy for all values of the
topological Hopf charge, including Hopf charge one.
On the other hand, it has been shown that in
(2+1) dimensions the energy of a $Q$-lump configuration is finite if
and only if the topological charge is at least two \cite{leese}.
\\
Another type of time-dependent configurations can be obtained in
the $CP^1$ model with a different kind of potential
\begin{equation}
\mathcal{L}=\frac{1}{4}{\cal L}_2 - V_{\rm II} \equiv
\frac{\partial_{\mu} u \partial^{\mu} \bar{u}}{(1+|u|^2)^2} -
\frac{\beta^2}{16} \left( \frac{1-|u|^2}{1+|u|^2} \right)^2.
\label{dyn model collaps}
\end{equation}
This potential also takes an elegant form if expressed by the
unit, vector field, namely
\begin{equation}
V_{\rm II}(\vec{n})= \frac{\beta^2}{16} (n^3)^2. \label{collaps
potential}
\end{equation}
In this case we obtain another massive modification of the free
$CP^1$ submodel with a real mass $M^2=\beta^2 /4$.
\\
One can check that now the time-dependent solutions are given by
\begin{equation}
u= C \left( \frac{1}{z}-1 \right)^{\pm \frac{m}{2}} e^{\pm
\frac{\beta}{2} t} e^{im(\phi_1+\phi_2)}, \label{dyn sol collaps}
\end{equation}
describing collapsing or exploding unknots.
\\
In spite of the fact that our time-dependent hopfions are not
sensitive to the radius of the sphere $R_0$, their energy is.
Thus, such solutions do not lead to finite energy configurations
in the limit $R_0 \rightarrow \infty$, that is, in three
dimensional Euclidean space.
\\
Finally, let us notice that, contrary to the pure $CP^1$ model, a
superposition of dynamical solutions derived for the submodels
(\ref{dyn submodel1}) and (\ref{dyn submodel2}) is no longer a
solution. This is due to nonlinearity of the constraint (\ref{dyn
submodel2}). However, we can obtain time-dependent multi-soliton
solutions if we assume that such a multi-soliton moves
collectively. That is to say, the general solution is
\begin{equation}
u = F(u_s) e^{\pm i\omega t}, \label{time sol general}
\end{equation}
where $u_s$ is an arbitrary static solution of the pure $CP^1$
model and $F$ is any (anti)\-holo\-morphic function.
%%%%%%%%%%%%%%%%%%%%%%%%%%%%%%%%%%%%%%%%%%%%%%%%%%%%%%%%%%%%%%%
\subsection{Generalized $CP^1$ models}
%%%%%%%%%%%%%%%%%%%%%%%%%%%%%%%%%%%%%%%%%%%%%%%%%%%%%%%%%%%%%%%
The obtained results may be easily generalized to more complicated
models. Namely, let us consider the following family of
Lagrangians
\begin{equation}
\mathcal{L}= \sigma (|u|^2)\; \frac{\partial_{\mu} u
\partial^{\mu} \bar{u}}{(1+|u|^2)^2} \label{cp1 model gen},
\end{equation}
where $\sigma (|u|^2)$ is any function of the modulus squared.
This family represents $CP^1$ models with a ``dielectric''
function $\sigma$. The equation of motion reads
\begin{equation}
\tilde{\sigma} \partial_{\mu} \partial^{\mu} u + \tilde{\sigma}'
\bar u \partial_{\mu} u  \partial^{\mu} u =0 ,\label{cp1 gen
eqmot2}
\end{equation}
where $\tilde{\sigma}\equiv \sigma / (1+|u|^2)^2$, and the prime
denotes the derivative w.r.t. the argument $|u|^2$. Thus, we get
that solutions of the submodel (\ref{cp1 submodel}) also obey
equation (\ref{cp1 gen eqmot2}) and, as a consequence, all
generalized models possess the same static solutions given by
(\ref{cp1 sol}).
\\
Similarly, time-dependent hopfions can be derived if we consider
the generalized $CP^1$ models with a potential
\begin{equation}
\mathcal{L}= \sigma (|u|^2) \left( \frac{\partial_{\mu} u
\partial^{\mu} \bar{u}}{(1+|u|^2)^2} - \frac{\beta^2}{4}
\frac{|u|^2  }{(1+|u|^2)^2} \right) .  \label{cp1 model gen time}
\end{equation}
The field equation is
\begin{equation}
\tilde{\sigma}' \bar u \left( (\partial_{\mu} u)^2
-\frac{\beta^2}{4}  \right) + \tilde{\sigma} (\partial^2_{\mu} u-
\frac{\beta^2}{4} u)=0. \label{cp1 gen osc eqmot1}
\end{equation}
Therefore, solutions of the dynamical subsystem of the $CP^1$
model (\ref{dyn submodel1}), (\ref{dyn submodel2}) satisfy
(\ref{cp1 gen osc eqmot1}) as well. That is to say, we have shown
that system (\ref{cp1 model gen time}) possesses stationary
hopfions
\begin{equation}
u= C \left( \frac{1}{z} -1 \right)^{\pm \frac{m}{2}} e^{\pm i
m(\phi_1 + \phi_2)} e^{\pm i \omega t} \label{cp1 gen osc sol}
\end{equation}
where the frequency obeys the relation $\omega^2 =\beta^2 /4$, as
before. As an interesting example let us mention a model with the
so-called "old baby Skyrme" potential
\begin{equation}
\mathcal{L}= \frac{(\partial \vec{n} )^2}{1+n^3} -
\frac{\beta^2}{16} (1-n^3). \label{old baby model}
\end{equation}
Analogously, one can construct Lagrangians which possess exact
collapsing/exploding time solitons (\ref{dyn sol collaps}),
\begin{equation}
\mathcal{L}= \sigma (|u|^2)\; \left( \frac{\partial_{\mu} u
\partial^{\mu} \bar{u}}{(1+|u|^2)^2} - \frac{\beta^2 }{4}
\left( \frac{1- |u|^2}{1+|u|^2} \right)^2 \right) .  \label{cp1 model gen
collaps}
\end{equation}
It is straightforward to obtain time-dependent multi-soliton
configurations.
%%%%%%%%%%%%%%%%%%%%%%%%%%%%%%%%%%%%%%%%%%%%%%%%%%%%%%%%%%%%%%%
\section{Other models}
%%%%%%%%%%%%%%%%%%%%%%%%%%%%%%%%%%%%%%%%%%%%%%%%%%%%%%%%%%%%%%%
%%%%%%%%%%%%%%%%%%%%%%%%%%%%%%%%%%%%%%%%%%%%%%%%%%%%%%%%%%%%%%%
\subsection{Hopfion with $Q_H=1$}
%%%%%%%%%%%%%%%%%%%%%%%%%%%%%%%%%%%%%%%%%%%%%%%%%%%%%%%%%%%%%%%
The aim of this section is to investigate a rather general family
of nonlinear sigma model on $S^3$. The unique restriction which we
assume is that the Lagrange density is any reasonable function of
the quantity
\begin{equation}
l=\frac{1}{4} {\cal L}_2 = \frac{\partial_{\mu} u
\partial^{\mu} \bar{u}}{(1+|u|^2)^2}. \label{dependence assump1}
\end{equation}
Thus, we will analyze the following models
\begin{equation}
\mathcal{L} =  \mathcal{L} \left( \frac{\partial_{\mu} u
\partial^{\mu} \bar{u}}{(1+|u|^2)^2} \right). \label{model q1}
\end{equation}
One well-known  member of that family is the Nicole model
$\mathcal{L}_{Ni}=4 l^{3/2}$. Of course, the pure $CP^1$ model
belongs to this family as well. However, as this case is rather
special, we have discussed it separately in the previous section.
\\
The equation of motion reads
\begin{equation}
\partial_{\mu} \left( \frac{\mathcal{L}'}{(1+|u|^2)^2} \partial^{\mu} u
\right) + \frac{2u}{(1+|u|^2)^3} \mathcal{L}' (\partial_{\mu} u
\partial^{\mu} \bar{u})=0, \label{model q1 eq mot0}
\end{equation}
or
\begin{equation}
\mathcal{L}' \partial_{\mu} \partial^{\mu} u +\mathcal{L}'' \,
(\partial_{\mu} l \, \partial^{\mu} u)- \frac{2\bar{u}}{1+|u|^2}
\mathcal{L}' (\partial_{\mu} u )^2=0, \label{model q1 eq mot1}
\end{equation}
where the prime denotes the differentiation with respect to $l$.
Analogously as for the pure $CP^1$ model it is possible to define
an integrable submodel
\begin{equation}
\partial^2_{\mu} u =0 \; \; \; (\partial_{\mu} u
)^2=0 \; \; \; \mbox{and} \;\;\; \partial_{\mu} l \,
\partial^{\mu} u=0. \label{model q1 submodel gen}
\end{equation}
As we see, such a subsystem is a restriction of the submodel for
the pure $CP^1$ model (\ref{cp1 submodel}), where the scalar field
must obey an additional equation. Therefore, a static soliton
solution in this submodel can be derived if we impose the
additional condition $\partial_{\mu} l \, \partial^{\mu} u=0$ on
the solutions of the pure $CP^1$ model obtained above.
\\
We calculate
\begin{equation}
l= \frac{1}{(1+f^2)^2} \frac{1}{R_0^2} \left(4z(1-z)f'^2+\frac{f^2
(m_1^2z+(1-z)m_2^2)}{z(1-z)} \right) \label{l eq1}
\end{equation}
or, if we put $m_1=m_2=m$,
\begin{equation}
l=\frac{1}{R_0^2} \frac{2f^2 m^2}{z(1-z)} \frac{1}{(1+f^2)^2}=
\frac{1}{R_0^2} \frac{2 m^2}{z(1-z)} \frac{\left( \frac{1}{z} -1
\right)^m}{\left[ 1+ \left( \frac{1}{z} -1 \right)^m \right]^2}.
\label{l eq2}
\end{equation}
We immediately notice that for $m=1$ we get $l=2/R_0^2=const.$ and
the additional condition is trivially obeyed. That means we have
constructed a solution of the submodel (\ref{model q1 submodel
gen}) i.e. a topological solution of the family of models
(\ref{model q1}) with unit Hopf index $Q_H=1$
\begin{equation}
u= \left( \frac{1}{z} -1 \right)^{\frac{1}{2}}
e^{i(\phi_1+\phi_2)} = \frac{X_3 +iX_4}{X_1 -iX_2} \label{model q1
sol}
\end{equation}
which is essentially (i.e., up to the reflection $X_2 \to -X_2$)
the standard Hopf fibration of $S^3$. Moreover, we can calculate
the energy of the soliton
\begin{equation}
E=  2 \pi^2 R_0^3 \mathcal{L} \left( \frac{2}{R_0^2} \right).
\label{model q1 energy}
\end{equation}
Observe that the  class of systems allowing for the hopfion
(\ref{model q1 sol}) is even largen than assumed in (\ref{model
q1}). In fact, all models depending additionally on a second
invariant
\begin{equation}
j=\frac{1}{8} {\cal L}_4 = \frac{(\partial_{\mu} u \partial^{\mu}
\bar{u})^2 - (\partial_{\mu} u
\partial^{\mu} u) (\partial_{\mu} \bar{u} \partial^{\mu} \bar{u})}{(1+|u|^2)^4}
\label{dependence assump2}
\end{equation}
also possess this hopfion with unit charge. This is due to the
trivial fact that the variable $j$ can be reduced to the variable
$l$ if the eikonal equation is satisfied. As this equation already
belongs to the submodel (\ref{model q1 submodel gen}) one can
conclude that (\ref{model q1 submodel gen}) is an integrable
submodel for all models of the form
\begin{equation}
\mathcal{L}=\mathcal{L} (l,j) \label{model q1 gen}
\end{equation}
with the nontrivial topological soliton (\ref{model q1 sol}). As a
consequence, as
\begin{equation}
\mathcal{L}_{FN}= 2 \left( \mu^2 \, l -\frac{1}{e^2} \, j \right).
\label{fn as q1}
\end{equation}
we are able to reproduce, within the generalized integrability,
the exact solution for the Faddeev--Niemi model on $S^3$ originally
found by R. Ward \cite{ward}.
%%%%%%%%%%%%%%%%%%%%%%%%%%%%%%%%%%%%%%%%%%%%%%%%%%%%%%%%%%%%%%%
\subsection{Stationary hopfions in the Faddeev--Niemi model}
%%%%%%%%%%%%%%%%%%%%%%%%%%%%%%%%%%%%%%%%%%%%%%%%%%%%%%%%%%%%%%%
The topic addressed in this subsection is the existence of a
time-dependent, stationary soliton in the Faddeev--Niemi model with
a potential term chosen as in (\ref{periodic potential}).
Therefore, we consider the Lagrangian ${\cal L} = \frac{1}{2}{\cal
L}_{\rm FN} - V_{\rm I}$, or, explicitly
\begin{equation}
{\cal L} = \mu^2 \frac{\partial_{\mu} u
\partial^{\mu} \bar{u}}{(1+|u|^2)^2} -\frac{1}{e^2} \frac{(\partial_{\mu} u
\partial^{\mu} \bar{u})^2
- (\partial_{\mu} u
\partial^{\mu} u) (\partial_{\mu} \bar{u} \partial^{\mu}
\bar{u})}{(1+|u|^2)^4} - \frac{\beta^2}{4}
\frac{|u|^2}{(1+|u|^2)^2} \label{fn oscillating}
\end{equation}
with the field equation
$$ \mu^2 (1+u\bar{u})^3 \partial_{\mu}^2 u - 2 \mu^2 (1+u\bar{u})^2 \bar{u}
u_{\mu}^2+$$
$$ \frac{\beta}{4} u(1-u\bar{u})(1+u\bar{u})^2 + \frac{4}{e^2} u [
(u^{\nu}\bar{u}_{\nu})^2 -u_{\mu}^2\bar{u}^2_{\nu}] - $$
\begin{equation}
\frac{2}{e^2} (1+u\bar{u}) [\bar{u}^{\mu \nu} u_{\mu} u_{\nu} -
u^{\mu \nu} \bar{u}_{\mu} u_{\nu} + u^{\mu} \bar{u}_{\mu}
\partial_{\nu}^2 u - u^2_{\mu} \partial_{\nu}^2 \bar{u}]=0.
\label{fn osc eq}
\end{equation}
Assuming $u=e^{i\omega t} v(\vec{r})$ results in
$$ -\mu^2(1+v\bar{v})^3 \triangle v +2\mu^2 (1+v\bar{v})^2 \bar{v}
(\nabla v)^2 + \left( \frac{\beta}{4} - \omega^2\mu^2 \right) v
(1-v\bar{v})(1+v\bar{v})^2+$$
$$ + \frac{4}{e^2} v [ -\omega^2 (v \nabla \bar{v} + \bar{v}
\nabla v)^2 +(\nabla v \cdot \nabla \bar{v})^2 -(\nabla v)^2
(\nabla \bar{v})^2] + $$
$$ \frac{2\omega^2}{e^2} v(1+v\bar{v})[ 2 \nabla v \cdot \nabla \bar{v}
+ \bar{v}
\triangle v + v \triangle \bar{v} ]-$$
\begin{equation}
\frac{2}{e^2} (1+v\bar{v})[
\bar{v}^{kj}v_kv_j-v^{kj}v_k\bar{v}_j+(\nabla v \cdot \nabla
\bar{v}) \triangle v - (\nabla v)^2 \triangle \bar{v}]=0.
\label{fn osc eq1}
\end{equation}
Now, we assume in addition that $(\nabla v)^2=0$ and $\triangle
v=0$. Therefore we get
$$ \left( \frac{\beta}{4} - \omega^2 \mu^2 \right)
v(1-v\bar{v})(1+v\bar{v})^2 + \frac{4\omega^2}{e^2} v (1-v\bar{v})
(\nabla v \cdot \nabla \bar{v}) +$$
\begin{equation}
\frac{4}{e^2} v (\nabla v \cdot \nabla \bar{v})^2 - \frac{2}{e^2}
(1+v\bar{v})(\nabla v \cdot \nabla \bar{v})^j v_j=0, \label{fn osc
eq2}
\end{equation}
where we used $\bar{v}^{kj}v_kv_j=(\bar{v}^kv_k)^j v_j$, which
holds because of the static eikonal equation. Next, we insert the
Ansatz $v=f(z)e^{i(m_1\phi_1+m_2\phi_2)}$, as in the previous
sections. Then the first line of equation (\ref{fn osc eq2})
becomes
\begin{equation}
v(1-f^2)\left[ \left( \frac{\beta}{4} - \omega^2\mu^2 \right)
(1+f^2)2+\frac{8\omega^2}{e^2R_0^2} f^2
\frac{zm_1^2+(1-z)m_2^2}{z(1-z)} \right], \label{fn osc part a1}
\end{equation}
where we used the relation
\begin{equation}
\nabla v \cdot \nabla \bar{v} = \frac{2}{R_0^2} f^2 \frac{zm_1^2
+(1-z)m_2^2}{z(1-z)},
\end{equation}
which follows from the static complex eikonal equation. In the
case when $m_1=m_2=1$, it leads to
\begin{equation}
v(1-f^2) \left[ \left( \frac{\beta^2}{4} -\omega^2\mu^2 \right)
(1+f^2)^2 + \frac{8\omega^2}{e^2R_0^2} f^2 \frac{1}{z(1-z)}
\right]. \label{fn osc part a2}
\end{equation}
Inserting the simplest Hopf map
$$ f=\left( \frac{1}{z} -1 \right)^{\frac{1}{2}},$$
this can be rewritten as
\begin{equation}
\frac{v(1-f^2)}{z} \left( \frac{\beta^2}{4} -\omega^2\mu^2
+\frac{8\omega^2}{e^2R_0^2} \right). \label{fn osc party a3}
\end{equation}
On the other hand, the second line of equation (\ref{fn osc eq2})
vanishes identically for the above profile function, as we know
already from Section 4.1 (see also \cite{ward}).
\\
As a result, the simplest Hopf map solves the equation of motion
(\ref{fn osc eq}) if the following dispersion relation is
satisfied
\begin{equation}
\frac{\beta^2}{4} -\omega^2\mu^2 +\frac{8\omega^2}{e^2R_0^2} =0.
\label{fn dispersion rel}
\end{equation}
Thus, if $\mu^2
> 8/e^2R_0^2$ we obtain a stationary solution of the Faddeev--Niemi
model with the new baby Skyrme potential
\begin{equation}
u= \left( \frac{1}{z}-1 \right)^{1/2} e^{\pm i (\phi_1+\phi_2)}
e^{\pm i\omega t} \label{fn osc sol}
\end{equation}
with the following frequency
\begin{equation}
\omega^2= \frac{\beta^2}{4} \frac{1}{\mu^2 - \frac{8}{e^2R_0^2}}.
\label{fn freq rel}
\end{equation}
The total energy reads
\begin{equation}
E=(2\pi)^2 \left[ \left( \mu^2R_0 +\frac{4}{e^2R_0} \right) +
\frac{\beta^2}{4} \left(
\frac{ R_0^3}{12} \frac{\mu^2e^2R_0^2 +4}{\mu^2e^2R_0^2 -8}
+\frac{ R_0^3}{6} \right) \right] . \label{energy qhopf FN
1}
\end{equation}
If the parameters of the model are chosen such that $\mu^2 <
8/e^2R_0^2$ then no stationary hopfion is found. However, for
these parameter choices one can obtain an stationary hopfion in a
slightly modified model. Namely, it is sufficient to take the
potential as in (\ref{collaps potential}).
\\
There is also a special case when $\mu^2 = 8/e^2R_0^2$. Now, in
order to fulfill Eq. (\ref{fn dispersion rel}), we must set $\beta
=0$ independently of the value of $\omega$. Therefore, now the
pure Faddeev--Niemi model without any potential term is
investigated. In other words, the solution (\ref{fn osc sol})
describes a stationary hopfion in the original Faddeev--Niemi
system with arbitrary frequency. Such a $Q$-hopfion possesses the
energy
\begin{equation}
E=(2\pi)^2 \frac{3}{2} \mu^2R_0 \left[ 1 + \frac{\omega^2
R_0^2}{12} \right]. \label{energy qhopf FN 2}
\end{equation}
As in the $CP^1$ model, $Q$-balls with unit topological charge are
finite energy configurations.
%%%%%%%%%%%%%%%%%%%%%%%%%%%%%%%%%%%%%%%%%%%%%%%%%%%%%%%%%%%%%%%
\subsection{Hopfions with $Q_H=m^2$}
%%%%%%%%%%%%%%%%%%%%%%%%%%%%%%%%%%%%%%%%%%%%%%%%%%%%%%%%%%%%%%%
It is possible to construct a slightly more complicated family of
models, analogously to \cite{were2}, which are solved by some of the other
eikonal knots with higher topological charges.
Concretely, we allow for the following dependence of the Lagrange
density
\begin{equation}
\mathcal{L}_{m}=  \mathcal{L} (l^{(m)})  \label{model qm^2}
\end{equation}
where
\begin{equation}
l^{(m)}=  \sigma^{(m)} (|u|^2) \cdot \,   \frac{\partial u
\partial \bar{u}}{(1+|u|^2)^2}  \label{l m def}
\end{equation}
and
\begin{equation}
\sigma^{(m)} (|u|^2)= \frac{(1+u\bar{u})^2}{u\bar{u}} \cdot
\frac{(u\bar{u})^{\frac{1}{m}}}{\left( 1+(u\bar{u})^{\frac{1}{m}}
\right)^2}. \label{sigma}
\end{equation}
For $m=1$, we just have the family of models investigated above.
The equation of motion takes the form
\begin{equation}
\partial_{\mu} \left[ \mathcal{L}_m' \frac{\sigma^{(m)}}{(1+|u|^2)^2}
\partial^{\mu} \right] - \mathcal{L}_m' \frac{\partial}{\partial \bar{u}}
\left[  \frac{\sigma^{(m)}}{(1+|u|^2)^2} \right]=0 \label{model qm
eqmot1}
\end{equation}
or equivalently
\begin{equation}
\frac{\sigma^{(m)} \mathcal{L}_m''}{(1+|u|^2)^2}  \partial_{\mu}
l^{(m)} \partial^{\mu} u +  \frac{\sigma^{(m)}
\mathcal{L}_m'}{(1+|u|^2)^2}
\partial_{\mu} \partial^{\mu} u + \mathcal{L}_m' \frac{\partial}{\partial
\bar{u}}  \left[ \frac{\sigma^{(m)}}{(1+|u|^2)^2} \right]
(\partial_{\mu} u)^2=0, \label{model qm eqmot2}
\end{equation}
where now prime denoted differentiation with respect to $l^{(m)}$.
Now, as before we can define a simpler submodel
\begin{equation}
\partial_{\mu} \partial^{\mu} u=0, \; \; \; (\partial_{\mu} u)^2=0
\;\;\; \mbox{and} \;\;\; \partial_{\mu} l^{(m)}
\partial^{\mu} u=0. \label{submodel qm}
\end{equation}
It consists of the standard pure $CP^1$ submodel part (the first
two formulas) and an addition condition. The knotted solutions of
the pure $CP^1$ model have been describe before see Eq. (\ref{cp1
sol}). Thus only the third equation in (\ref{submodel qm}) needs
to be solved. For this purpose we insert the static knotted
solutions of the free $CP^1$ model,
$$ u_k= \left( \frac{1}{z} -1 \right)^{ \frac{k}{2}}
e^{ik(\phi_1+\phi_2)}, $$ into (\ref{l m def}). Then, after a
simple calculation one observes that $l^{(m)}$ is constant if and
only if $m=k$. In other words, each family of models with fixed
value of the parameter $m = 1,2,3...$, i.e., based on the variable
$l^{(m)}$, possesses a topological soliton solution
\begin{equation}
u_m= \left( \frac{1}{z} -1 \right)^{\frac{m}{2}}
e^{im(\phi_1+\phi_2)} \label{nicole gen sol}
\end{equation}
with the Hopf index $Q_H=m^2$.
\\
In analogy to the $Q_H=1$ solution such solitons can be also found
in all possible models of the form
\begin{equation}
\mathcal{L}_m= \mathcal{L} (l^{(m)},j^{(m)}) \label{def fn gen}
\end{equation}
based also on the additional variable
\begin{equation}
j^{(m)}= ( \sigma^{(m)})^2  \frac{(\partial_{\mu} u
\partial^{\mu} \bar{u})^2 - (\partial_{\mu} u
\partial^{\mu} u) (\partial_{\mu} \bar{u}
\partial^{\mu} \bar{u})}{(1+|u|^2)^4}. \label{fn gen}
\end{equation}
As a result, we find the interesting fact that there exists a
family of modified Fa\-dd\-eev-Niemi models which possess exact
topological knotted solitons. Namely, the models defined by the
following Lagrange density
\begin{equation}
\mathcal{L}_{FN}^{m}= \mu^2 \sigma^{(m)}
 \frac{\partial_{\mu} u
\partial^{\mu} \bar{u}}{(1+|u|^2)^2}-\frac{1}{e^2} (\sigma^{(m)})^2
\frac{(\partial_{\mu} u
\partial^{\mu} \bar{u})^2 - (\partial_{\mu} u
\partial^{\mu} u) (\partial_{\mu} \bar{u} \partial^{\mu} \bar{u})}{(1+|u|^2)^4}
\end{equation}
have the soliton solutions (\ref{nicole gen sol}).
%%%%%%%%%%%%%%%%%%%%%%%%%%%%%%%%%%%%%%%%%%%%%%%%%%%%%%%%%%%%%%%
\section{Summary and discussion}
%%%%%%%%%%%%%%%%%%%%%%%%%%%%%%%%%%%%%%%%%%%%%%%%%%%%%%%%%%%%%%%
In the present paper, sigma-model type field theories with a field
contents parametrized by the unit, three-component, vector field
living on $S^3 \times \RR$ space-time have been investigated.
There are two reasons for choosing such a physical space-time.
First of all, it stabilizes, at least for some models and for some
value of the
parameters, the obtained solitons by introducing a scale
parameter, i.e., the radius of the sphere $R_0$. Moreover, it also
enables us for a rather big family of models to obtain at least
some solutions in exact form. Specifically, we obtain
solutions in many cases where the corresponding theories on
space-time $\RR^3 \times \RR$ either do not have solutions (like,
e.g., for the $CP(1)$ model) or where there are no solutions known
analytically (like, e.g., for the Faddeev--Niemi model). Whereas
the first issue can be explained through Derrick's theorem, which
does not hold for the three sphere, the second one can be related
to the different isometry groups of $\RR^3$ and $S^3$,
respectively. Indeed, the isometry group of $S^3$ is $SO(4)$ which
has rank 2. Therefore, there exist two commuting vector fields
(generators of isometries) which can be chosen to be ${\bf v}_1
=\partial_{\phi_1}$ and ${\bf v}_2 =\partial_{\phi_2}$. These are
symmetry generators for all theories where the Lagrangian is a
scalar, therefore the ansatz (\ref{anzatz}) is compatible with the
e.o.m. for all such theories and reduces the static e.o.m. to a
nonlinear ODE. On the other hand, on $\RR^3$ the isometry is only
$SO(3)$ with rank one (forgetting the irrelevant translations). To
get a second commuting vector field (e.g., the angles $\xi$ and
$\varphi$ of the toroidal coordinates) one has to extend the
symmetry of the model under consideration (e.g., by choosing
theories with conformally invariant static e.o.m., as for the
Nicole and AFZ models; for a detailed account we refer to
\cite{AFZ-sym}). In more general cases like, e.g., for the
Faddeev--Niemi model, only a symmetry reduction to two independent
variables is possible, and the resulting non-linear PDE in two
variables is still too complicated to be solved analytically.

\subsection{Stability}
Next we want to discuss the issue of stability of our static solutions.
For this purpose it is useful to briefly recall the situation in flat space
$\RR^3$. For actions which are homogeneous in the degree of derivatives,
a scaling instability is present and prevents the existence of soliton 
solutions (static solutions). This is the contents of Derrick's theorem. 
The instability is due to ultra violet (UV) collapse of field configurations
when the homogeneous degree of derivatives is less than three, and due to
infra red (IR) collapse for a homogeneous degree greater than three.
For a homogeneous degree exactly equal to three the energy of a static
field configuration is invariant under scaling, and static solutions may
exist. In addition, the group of base space symmetries of the static e.o.m.
is enhanced (e.g., conformal symmetries instead of isometries). This is
exactly what happens, e.g., in the Nicole and AFZ models.
If, on the other hand, the theory consists of a sum of terms with
different degrees of derivatives such that at least one has degree less
than three, and at least one has degree greater than three, then these
two terms scale oppositely under scale transformations, and static
solutions may exist. Further, the model is not scale invariant, therefore
solitons have a typical "size" as an intrinsic property. This is the case,
e.g., for the Faddeev--Niemi model.  

Now let us discuss the analogous situation on the sphere $S^3$.
On the three-sphere an IR collapse is no longer possible. A field 
configuration which obeys a non-trivial boundary condition (i.e., which
has a non-zero topological index) will always contribute some
nonzero values of derivatives over some finite subvolumes of the 
entire $S^3$. On the other hand, an UV collapse (shrinking of field
configurations) is still possible. Therefore, we expect that theories
with actions which are homogeneous with less than three derivatives
will not have genuine solitons - i.e., static solutions which are 
absolute minima of the energy within a sector with fixed topological 
charge. On the other hand, for models which contain at least one term 
with degree in derivatives greater than three we expect stable solutions,
i.e., genuine solitons. 

We want to investigate the issue of stability more
closely for the simplest Hopf map  (\ref{model q1 sol}), which solves
most of the theories we have studied in this paper. The
stability issue of this field configuration has already been investigated 
in Reference \cite{ward} for the Faddeev--Niemi model, so we can make
use of these results. In \cite{ward} a one-parameter family of fields
$u_\lambda$ has been constructed, where $\lambda =1$ just gives the 
standard Hopf map. The energy density of the standard Hopf map is 
constant both for the quadratic lagrangian ${\cal L}_2$ (the $CP^1$
term) and for the quartic lagrangian ${\cal L}_4$. Further, both energy
densities become peaked for very large or very small values of the
parameter $\lambda$ (around the north pole or south pole of the $S^3$,
respectively). The energy of the quadratic ($CP^1$) term for the 
one-parameter family $u_\lambda$ has a maximum at $\lambda =1$. For
very small or very large values of $\lambda$ the energy approaches zero.
However, the limiting field configurations for $\lambda =0$ or $\lambda
= \infty$ cannot be attained, because they are trivial and do not belong
to the sector with Hopf index one. Therefore, there does not exist a
genuine soliton in the sector with Hopf index one for the $CP^1$ model.
The solution for $\lambda =1$ is a saddle point solution rather than a
minimum. For the quartic term, the energy has a minimum at $\lambda =1$.
Further, the energy tends to infinity in the limits $\lambda \to 0$ and
$\lambda \to \infty$. This supports the conjecture that the standard
Hopf map is a genuine soliton (minimizer of the sector with Hopf index one)
for the quartic model (although there does not seem to exist a rigorous proof
up to now). For the case of the Faddeev--Niemi model ${\cal L}_{\rm FN}
= {\cal L}_2 - {\cal L}_4$ (here we ignore constants) we just briefly
repeat the discussion of Reference \cite{ward}. For sufficiently small
radius $R_0$ of the three-sphere the energy of the quartic term dominates  
(behaving like $1/R_0$), and the energy is minimized for $\lambda =1$.
So probably the standard Hopf map is a true minimum. For large values
of the sphere radius the energy of the quadratic term dominates (behaving like
$R_0$), and the standard Hopf map is just a saddle point. However, now
complete UV collapse is not possible (this would render the energy of the
quartic term infinite). Instead the energy is minimized for some finite 
$\lambda_0 \ge 1$ (or, equivalently, for its inverse $1/\lambda_0$) with
the energy density localized around the north pole (or south pole) of the
$S^3$. 
For larger values of $R_0$ the localisation becomes more pronounced (i.e.,
$\lambda_0$ becomes larger). So a true soliton probably exists for the
Faddeev--Niemi model even for large values of the sphere radius, but it is
no longer the standard Hopf map with its energy density evenly distributed 
over the whole $S^3$. We expect this generic pattern of stability also to hold
for higher Hopf index. The generalization of the above discussion of stability
to the other models studied in this paper is straight forward.

Finally, let us just mention that the question of stability is more involved 
for the stationary solutions (Q-balls). Firstly, stability is no longer
related to the minimization of the energy and, secondly, the presence
of further nontrivial conserved charges (like the Noether charge in
Section 3.2) complicates the analysis and tends to make
solutions more stable. 
A detailed discussion of that issue is beyond the scope of
this article.

\subsection{Summary of results}
Firstly, static  knotted configurations solving the complex
eikonal equation have been derived. They are the $S^3$
counterparts of the eikonal knots on $\RR^3$ and, therefore,
describe linked torus knots with arbitrary value of the
topological charge. The problem whether non-torus knots,
represented for instance by the figure-eight knot, can also be
found for the eikonal equation is still an open question.
Unfortunately, our method does not allow us construct such knots.
In addition, time-dependent knots (stationary or
exploding/collapsing ones) have been constructed.
\\
Secondly, we have shown that eikonal knots with 
Hopf index $Q_H=\pm m^2$, where $m \in Z$, are solutions
of the pure $CP^1$ model on $S^3$. The
energies of these solutions can be related to
their topological charges. Concretely, the energy is proportional
to the square root of the charge. Stability analysis shows that these
solutions are not stable, i.e., they are not true solitons. 
Instead, they are saddle point solutions.
A family of exact stationary
solutions has been obtained, as well, for the $CP^1$ model with the
"new baby Skyrme" potential term. Their frequencies are strictly
determined and do not depend on the topology of the solutions
(value of the parameters $m_1$ and $m_2$). Moreover, a slight
modification of the potential gives collapsing/exploding
solutions.
\\
Thirdly, in a very large class of models a static hopfion with
unit Hopf index (the standard Hopf fibration) has been found. In
the case of the Faddeev--Niemi model, we reproduced a solution
already obtained by R. Ward, \cite{ward}. In addition, a
stationary generalization of the soliton has been derived for the
Faddeev--Niemi model with the "new baby Skyrme" potential. Its
frequency is determined by the parameters of the model. This may
be of some interest in the context of the effective model for the
low energy quantum gluodynamics. Namely, the Faddeev--Niemi model
spontaneously breaks the global $O(3)$ symmetry and, as a
consequence, two massless Goldstone bosons appear. To get rid of
such nonphysical excitations one has to improve the model and add
a symmetry breaking term \cite{FaNi3} (see also \cite{wipf}). The
most obvious way to accomplish this is to introduce a potential. For
the special case when the parameters obey $\mu^2 = 8/e^2R_0^2$ a
stationary hopfion with $Q_H=1$ has been found in the pure
Faddeev--Niemi model, where the frequency may take on arbitrary
values.
\\
Finally we have proved that also more complicated static hopfions
with higher values of the topological charge can be obtained in
modified models. The modification is given by the so-called
dielectric function.
\\
There are several directions in which our work can be continued.
One could, for example, try to derive static soliton solutions in the
models with the new baby Skyrme potential added, and compare them
with the stationary solutions.
\\
On the other hand, one could study the issue of quantization of
the obtained Hopf solitons \cite{quantum}. Also the relevance of the
saddle point solutions of the $CP^1$ model for its subsequent
quantization would be worth investigating. 
\\
Finally, we hope that the results presented here lead to some
further insight into general properties of theories with knotted
solitons and may, in this respect, also help in understanding the
corresponding theories in standard Minkowski space-time.
\\ \\ \\
{\large\bf Acknowledgement:} \\
This research was partly supported by MCyT(Spain) and FEDER
(FPA2005-01963), Incentivos from Xunta de Galicia and the EC
network "EUCLID". Further, CA acknowledges support from the
Austrian START award project FWF-Y-137-TEC and from the  FWF
project P161 05 NO 5 of N.J. Mauser. AW gratefully acknowledges
support from the Polish Ministry of Education and Science.

\end{document}